\documentclass[twocolumn,prd,floatfix,amsmath,nofootinbib,amssymb,floatfix]{revtex4}
\usepackage{graphicx,color,dcolumn,booktabs,bm}
\usepackage{longtable,lscape}
\usepackage{txfonts}
\usepackage{overpic}
\usepackage{amssymb}
\usepackage{indentfirst}
\usepackage{feynmf}
\usepackage{slashed}
\usepackage{cases}
\usepackage{multirow}
\usepackage[colorlinks,
            citecolor=blue,
            anchorcolor=red,
            menucolor=red,
            linkcolor=red,
            filecolor=red,
            runcolor=red,
            urlcolor=blue,
            frenchlinks=red]{hyperref}
\usepackage{epstopdf}
\usepackage{bm}
\begin{document}
\title{Predictions for feed-down enhancements at the $\Lambda_c \bar{D}$ and $\Lambda_c \bar{D}^*$ thresholds via the triangle and box singularities}
\author{Ming-Xiao Duan$^{1}$}\email{duanmx@ihep.ac.cn}
\author{Lin Qiu$^{1,2}$}\email{qiulin@ihep.ac.cn}
\author{Xi-Zhe Ling$^{1}$}\email{lingxz@ihep.ac.cn}
\author{Qiang Zhao$^{1,2}$}\email{zhaoq@ihep.ac.cn}

\affiliation{
$^1$ Institute of High Energy Physics,~Chinese Academy of Sciences, Beijing 100049,~P.R.~China and\\
$^2$ University of Chinese Academy of Sciences,~Beijing 100049,~P.R.~China}

\begin{abstract}
We demonstrate that triangle singularity (TS) and box singularity (BS) mechanisms can produce unique narrow enhancements at the $\Lambda_c\bar{D}$ and $\Lambda_c\bar{D}^*$ thresholds in the invariant mass spectra of $J/\psi p$ and $J/\psi p\pi$, respectively.  Taking into account that such mechanisms only depend on the initial $\Sigma_c^{(*)}\bar{D}^{(*)}$ interactions near threshold within the TS or BS kinematic regimes, the $\Lambda_c\bar{D}$ and $\Lambda_c\bar{D}^*$ threshold enhancements can be regarded as a feed-down phenomenon originated from both the heavier pentaquark decays and the $\Sigma_c^{(*)}\bar{D}^{(*)}$ scatterings from the continuum. A search for these structures in the $J/\psi p$ and $J/\psi p\pi$ spectra in both exclusive and semi-inclusive processes will provide a smoking-gun evidence for the hadronic molecule nature of those observed pentaquarks and clarify the role played by the TS and BS in the near-threshold dynamics.

\end{abstract}

\maketitle


{\it Introduction} \
The observation of hidden charm pentaquarks by the LHCb Collaboration in 2015~\cite{LHCb:2015yax} provided a strong evidence for QCD exotics in the baryon sector and has initiated dramatic activities in both experiment and theory to understand their nature (see review articles \cite{Zhao:2016akg, Guo:2017jvc, Chen:2016qju,Olsen:2017bmm,Esposito:2016noz,Lebed:2016hpi,Brambilla:2019esw,Liu:2019zoy,Ali:2017jda,Karliner:2017qhf,Chen:2022asf} for details). Based on the Run-1 data sample, two states, $P_c(4380)$ and $P_c(4450)$, were observed in the $J/\psi p$ invariant mass spectrum in $\Lambda_b^0\to J/\psi pK^-$~\cite{LHCb:2015yax}. Since their masses were close to the threshold of $\Sigma_c^{(*)}\bar{D}^{(*)}$ channel, they were interpreted as hadronic molecules formed by the $\Sigma_c^{(*)}\bar{D}^{(*)}$ near-threshold interactions~\cite{Chen:2015loa, Chen:2015moa, Karliner:2015ina, Roca:2015dva, He:2015cea, Lu:2016nnt, Shen:2016tzq, Yamaguchi:2016ote, Lin:2017mtz, Yamaguchi:2017zmn, Wang:2018waa}. With the Run-2 data samples, the reanalysis of $\Lambda_b^0\to J/\psi pK^-$ exposes more detailed structures in the $J/\psi p$ invariant mass spectrum~\cite{LHCb:2019kea}. $P_c(4450)$ splits into two states, $P_c(4440)$ and $P_c(4457)$, which fit well the $J^P=1/2^-$ and $3/2^-$ doublet formed by $\Sigma_c\bar{D}^*$, and a new state $P_c(4312)$, which fits the $\Sigma_c\bar{D}$ molecular state, is identified~\cite{Chen:2019bip, Chen:2019asm, Liu:2019tjn, He:2019ify, Guo:2019kdc, Xiao:2019mvs, Xiao:2019aya, Meng:2019ilv, Yamaguchi:2019seo, Du:2019pij, Wang:2019ato}. Meanwhile, signals for $P_c(4380)$, formerly assigned as the $\Sigma_c^*\bar{D}$ molecule in the Run-1 analysis, become obscure in the Run-2 analysis. It turns out to be broad and can be even absorbed by the background \cite{LHCb:2019kea}. It should be noted that the quantum numbers determined by the Run-2 analysis are more consistent with the molecular scenario based on the relative $S$-wave interactions and also consistent with the early theoretical predictions~\cite{Wu:2010vk,Yang:2011wz,Wang:2011rga,Wu:2012md}.

A key issue clarified by the Run-2 analysis is that these hidden charmonium pentaquark candidates couple to the nearby $\Sigma_c\bar{D}^{(*)}$ in a relative $S$-wave. This makes the hadronic molecule scenario a natural interpretation for their internal structures and provide important insights into the threshold dynamics. Note that these observed $P_c$ states are correlated with the $\Sigma_c^{(*)}\bar{D}^{(*)}$ thresholds instead of $\Lambda_c^{(*)}\bar{D}^{(*)}$. It is also interesting to observe that this feature seems to be consistent with the one boson exchange (OBE) dynamics between the charmed meson and baryon. Namely, it manifests to be attractive between $\Sigma_c^{(*)}$ and $\bar{D}^{(*)}$, but appears to be repulsive between $\Lambda_c^{(*)}$ and $\bar{D}^{(*)}$~\cite{Wang:2011rga, Yang:2011wz,Chen:2019bip, Chen:2019asm, Liu:2019tjn, He:2019ify, Guo:2019kdc, Xiao:2019mvs, Xiao:2019aya, Meng:2019ilv, Yamaguchi:2019seo, Du:2019pij, Wang:2019ato}. This can explain that no obvious enhancements have been seen at the thresholds of $\Lambda_c\bar{D}$ and $\Lambda_c\bar{D}^*$ in exclusive decays of $\Lambda_b^0\to J/\psi pK^-$~\cite{LHCb:2019kea}.

Another important threshold dynamics to be exposed by the pentaquark production in the molecular picture is that the $\Sigma_c^{(*)}\bar{D}^{(*)}$ configurations in the $\Lambda_b$ decays would be either produced via the color-suppressed process, or via their subleading couplings to $\Lambda_c^{(*)}\bar{D}^{(*)}$ through the triangle singularity (TS)~\cite{Landau:1959fi,Cutkosky:1960sp} enhanced transitions as pointed out in Refs.~\cite{Liu:2015fea,Guo:2015umn}. Notice that the $\Lambda_c\bar{D}^{(*)}$ configurations can be accessed by both the direct emission processes and TS enhanced transitions. The absence of enhancements around the thresholds of $[\Lambda_c\bar{D}](4150)$ and $[\Lambda_c\bar{D}^*](4290)$ seems to confirm that those observed hadronic molecules states are genuine states driven by the $\Sigma_c^{(*)}\bar{D}^{(*)}$ interactions, while the repulsive nature of $\Lambda_c^{(*)}\bar{D}^{(*)}$ would not produce apparent threshold structures at leading order in the exclusive decays of $\Lambda_b$. In Ref.~\cite{Wang:2015jsa} it was proposed to search for these heavy pentaquark states in $J/\psi$ photoproduction where the cross sections at large scattering angles should serve as direct probe for the production of the pentaquarks as genuine states. Although these hidden charm pentaquark states are searched at GlueX based on their 12 GeV data, the present statistics do not allow a conclusion for their existence at this moment~\cite{GlueX:2019mkq}.

Apparently, the confirmation or refusal of $P_c(4380)$ will have a strong impact on our understanding of the nature of these hidden charm pentaquarks. In this work, based on the hadronic molecule picture, we identify the crucial role played by the TS and BS in the pentaquark three-body and four-body decays into $J/\psi p \pi$ and $J/\psi p\pi\pi$, respectively. It will strongly enhance the width of $P_c(4380)$ due to the special kinematic in $P_c(4380)\to J/\psi p\pi$, and naturally accommodate it into the hadronic molecule picture. As a smoking-gun evidence for this mechanism, we predict two feed-down enhancements at the $\Lambda_c\bar{D}$ and $\Lambda_c\bar{D}^*$ thresholds in the $J/\psi p$ and $J/\psi p\pi$ invariant mass spectra, respectively. We will discuss that high-statistics data for exclusive processes or semi-inclusive channels of $J/\psi\pi$ and $J/\psi\pi\pi$ will be ideal for the search for these two threshold structures.

\begin{center}
\begin{figure}[htbp]
\includegraphics[width=7cm,keepaspectratio]{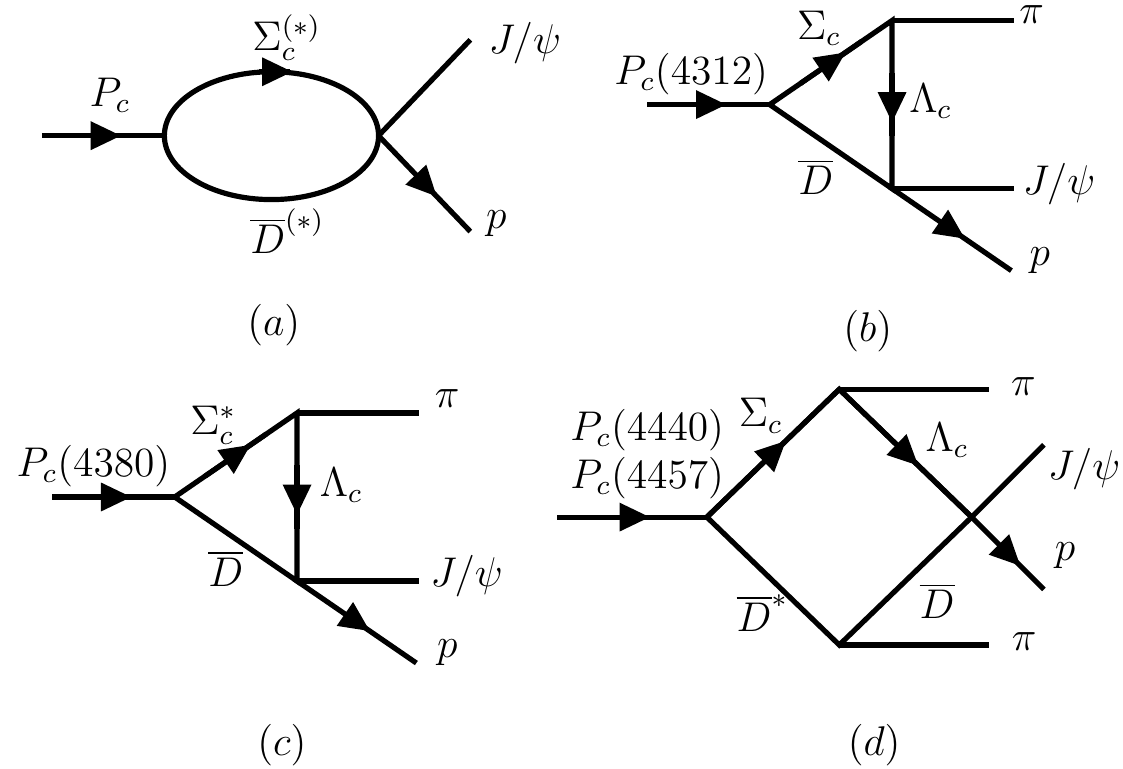}
\caption{Feynman diagrams illustrating the $P_c$ states decay into $J/\psi p$, $J/\psi p\pi$, and $J/\psi p\pi\pi$. Figures (b)-(d) manifest the TS and BS mechanisms that can feed down to the semi-inclusive $J/\psi p$ and $J/\psi p\pi$ spectra, respectively.}
\label{f1}
\end{figure}
\end{center}

{\it Formalism} \
Given the productions of these $P_c$ states via various processes, their decays into $J/\psi p$ can be via a direct two-body decays into $J/\psi p$, three-body decays into $J/\psi p\pi$, or four-body decays into $J/\psi p\pi\pi$. An illustration of these transitions is presented in Fig.~\ref{f1}. Note that in both Fig.~\ref{f1} (b) and (c) the transition satisfy the TS condition, i.e. all the internal particles can approach their on-shell condition simultaneously. As the leading singularity in the loop functions, it may provide non-negligible partial widths for these $P_c$ states via their $J/\psi p\pi$ three-body decays. In comparison with the two-body decay amplitude which counts an order of $v$, where $v$ is the typical non-relativistic on-shell velocity of the intermediate heavy hadrons, the triangle loop of Fig.~\ref{f1} (a) and (b) count $(v^5/v^6)v\sim {\cal O}(1)$.

The transitions of Fig.~\ref{f1} (d) fulfill the so-called ``box singularity" (BS), where all the internal particles can also approach their on-shell conditions simultaneously. The amplitudes of Fig.~\ref{f1} (d) count $(v^5/v^8)v^2\sim 1/v$ which appears to be more leading than (b) and (c). However, taking into account the limited phase space, contributions from $P_c(4440)$ and $P_c(4457)\to J/\psi p\pi\pi$ will be suppressed compared to those three-body decays.

In the feed-down scenario the semi-inclusive spectra of $J/\psi p$ or $J/\psi p\pi$ will be measured. It means that all the transitions in Fig.~\ref{f1} will contribute and sum from the $J/\psi p$ or $J/\psi p\pi$ threshold to higher energies. Due to the triangle singularity, we would expect the threshold enhancement at the mass of $\Lambda_c\bar{D}$ in the $J/\psi p$ invariant mass spectrum, and due to the box singularity, we would expect the threshold enhancements at the thresholds of $\Lambda_c\bar{D}^*$ and $\Sigma_c\bar{D}$ in the $J/\psi p\pi$ invariant mass spectrum.

\begin{center}
\begin{figure}[htbp]
\includegraphics[width=6.cm,keepaspectratio]{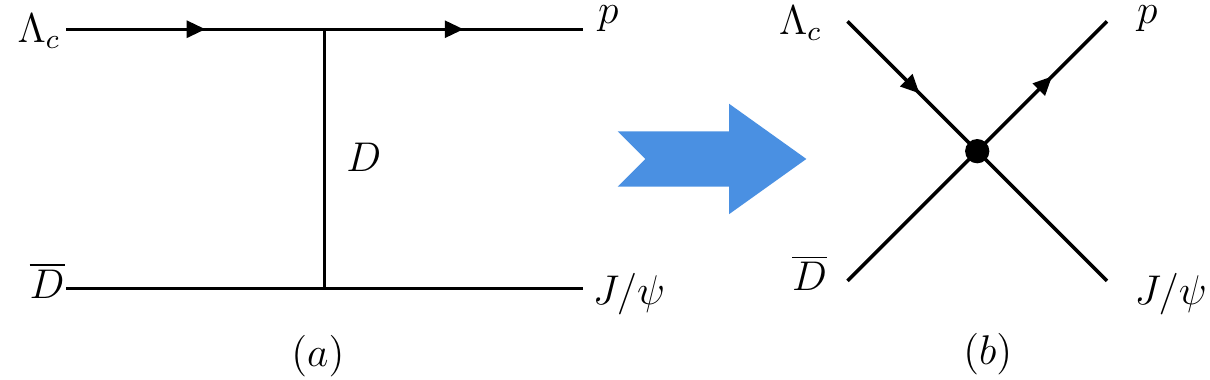}
\caption{The near-threshold interaction between $\Lambda_c\bar{D}$ and $J/\psi p$ is described by a $t$-channel $D$ meson exchange from which the effective contact coupling $g_x$ is extracted. }
\label{f2}
\end{figure}
\end{center}

To proceed, we employ an effective Lagrangian approach to evaluate the feed-down mechanisms. The $S$-wave interactions between $P_c$ and $\Sigma_c^{(*)} \bar{D}^{(*)}$ can be expressed as~\cite{Zou:2002yy}
\begin{equation}
\begin{split}
\mathcal{L}_{P_c(4312)\Sigma_c\bar{D}}&=g_{P_c(4312)}\bar{D}^\dag\bar{\Sigma}_cP_c,\\
\mathcal{L}_{P_c(4380)\Sigma_c^*\bar{D}}&=g_{P_c(4380)}\bar{D}^\dag\bar{\Sigma}^*_c\cdot P_c,\\
\mathcal{L}_{P_c(4440)\Sigma_c\bar{D}^*}&=g_{P_c(4440)}\bar{D}^{*\mu\dag}\bar{\Sigma}_c\gamma_5\tilde{\gamma}_\mu(p_A)P_c,\\
\mathcal{L}_{P_c(4457)\Sigma_c\bar{D}^*}&=g_{P_c(4457)}\bar{\Sigma}_c\bar{D}^{*\dag}\cdot P_c,\\
\end{split}
\end{equation}
where $\bar{\Sigma}_c^{(*)}$ and $\bar{D}^{(*)\dag}$ denote the field operators for $\Sigma_c^*$ and $\bar{D}^{(*)}$, respectively, and $P_c$ denotes the spinor field of $P_c$ states; $g_{P_c}$ are the coupling constants between $P_c$ and corresponding $\Sigma_c^{(*)} \bar{D}^{(*)}$ channels. They are determined in the molecular scenario by reproducing the mass and width of the $P_c$ states.
We adopt the coupling constants $g_{P_c(4312)}=2.16$, $g_{P_c(4380)}=2.19$, $g_{P_c(4440)}=2.6$, and $g_{P_c(4457)}=1.7$ determined from Ref.~\cite{Xie:2022hhv}. $\tilde{\gamma}_\mu(p_A)$ is a momentum dependant function, which is defined as $\tilde{\gamma}^\mu(p_A)=(-g^{\mu\nu}+p_A^{\mu}p_{A}^{\nu}/p_A^2)\gamma_\nu$~\cite{Zou:2002yy}. The projection operator for the spin-3/2 particle can be written as
\begin{eqnarray}
\sum u^\nu\bar{u}^\alpha&=&(\slashed{p}+m)[-g^{\nu\alpha}+\frac{1}{3}\gamma^\nu\gamma^\alpha\nonumber\\
&&+\frac{1}{3m}(\gamma^\nu p^{\alpha}-\gamma^{\alpha}p^{\nu})+\frac{2}{3m^2}p^\nu p^\alpha],
\end{eqnarray}
with which the propagator of $\Sigma_c^*$ in Fig.~\ref{f1} (b) can be obtained. Lagrangians for other vertices in Fig.~\ref{f1} are collected as follows:
\begin{eqnarray}\label{e1}
\mathcal{L}_{\Lambda_c\bar{D}J/\psi p}&=&ig_x\psi^{\mu}\bar{N}\gamma_5\gamma_\mu \Lambda_c\bar{D},\nonumber\\
\mathcal{L}_{\Sigma_c\Lambda_c\pi}&=&g_{\Sigma_c\Lambda_c\pi}\bar{\Lambda}_c\partial^\mu\pi\gamma_5\gamma_\mu \Sigma_c,\nonumber\\
\mathcal{L}_{\Sigma^*_c\Lambda_c\pi}&=&g_{\Sigma_c^*\Lambda_c\pi}\bar{\Lambda}_c\partial^{\mu}\pi\Sigma_{c\mu}^*,\nonumber\\
\mathcal{L}_{D^*D\pi}&=&ig_{D^*D\pi}D_\mu^*\bar{D}\partial^\mu\pi \ ,
\end{eqnarray}
where the coupling constants $g_{\Sigma_c\Lambda_c\pi}$, $g_{\Sigma_c^*\Lambda_c\pi}$, and $g_{D^*D\pi}$ are determined by the partial decay widths of the corresponding processes~\cite{ParticleDataGroup:2022pth}. Coupling $g_x$ describes the contact interaction for the $\Lambda_c \bar{D} J/\psi p$ vertex. It can be estimated by a $t$-channel $D$ meson exchange near threshold as shown in Fig.~\ref{f2}.

The Lagrangians involved in the determination of $g_x$ in Fig.~\ref{f2} (a) are
\begin{equation}
\begin{split}
\mathcal{L}_{\Lambda_cDp}&=ig_{DN\Lambda_c}D^\dag\bar{N}\gamma_5\Lambda_c,\\
\mathcal{L}_{J/\psi D\bar{D}}&=g_{J/\psi D\bar{D}}(D\partial_\mu\bar{D}-\partial_\mu D\bar{D})\psi^{\mu},\\
\end{split}
\end{equation}
where the coupling constant $g_{DN\Lambda_c}=13.2$ is estimated the same as the coupling of $KN\Lambda$ by employing the $SU(4)$ symmetry~\cite{Rijken:1998yy, Stoks:1999bz, Oh:2006hm, Liu:2010um}; Coupling $g_{J/\psi D\bar{D}}=2g_2\sqrt{m_{J/\psi}m_Dm_{\bar{D}}}$ with $g_2=\sqrt{m_{\psi}}/(2m_Df_{J/\psi})$ and $f_{J/\psi}=0.405~{\rm GeV}$~\cite{Colangelo:2003sa} is adopted from the heavy quark symmetry relation. As illustrated in Fig.~\ref{f2}, the relationship $|\mathcal{M}^{(a)}|^2=|\mathcal{M}^{(b)}|^2$ is employed, which has a concrete form like:
\begin{eqnarray}
g_x^2&=&\frac{g_{DN\Lambda_c}^2g_{J/\psi D\bar{D}}^2 |\bar{u}_p\gamma_5u_{\Lambda_c}(p_{\bar{D}}-q_{D})_{\mu}\frac{1}{q_{D}^2-m_D^2}\varepsilon^{*\mu}_{J/\psi}|^2}{|\bar{u}_{p}\gamma_5\gamma_\mu u_{\Lambda_c}|^2},
\end{eqnarray}
which yields $|g_x|=3.0~{\rm GeV}^{-1}$ at the threshold of $\Lambda_c$ and $\bar{D}$.

With the above Lagrangians, the triangle and box loop amplitudes in Fig.~\ref{f1} can be expressed as
\begin{equation}
\begin{split}
&\mathcal{M}^{P_c(4312)\to J/\psi p\pi}\\
&=\int\frac{dp^4_{\Lambda_c}}{(2\pi)^4}\bar{u}_p\gamma_5\gamma_\mu \varepsilon_{J/\psi}^{*\mu}
(\slashed{p}_{\Lambda_c}+m_{\Lambda_c})\gamma_5\gamma_\nu p_\pi^\mu(\slashed{p}_{\Sigma_c}+m_{\Sigma_c})u_{P_c}\\
&\quad\times\frac{g_{P_c(4312)}g_{\Sigma_c\Lambda_c\pi}g_x}{(p_{\Sigma_c}^2-m_{\Sigma_c}^2)(p_{D}^2-m_D^2)(p_{\Lambda_c}^2-m_{\Lambda_c}^2)},\\
&\mathcal{M}^{P_c(4380)\to J/\psi p\pi}\\
&=\int\frac{d^4p_{\Lambda_c}}{(2\pi)^4}\bar{u}_p\gamma_5\gamma_\mu \varepsilon_{J/\psi}^{*\mu}
(\slashed{p}_{\Lambda_c}+m_{\Lambda_c})p_{\pi\nu}(\slashed{p}_{\Sigma_c}+m_{\Sigma_c})\\
&\quad\times[-g^{\nu\alpha}+\frac{1}{3}\gamma^\nu\gamma^\alpha+\frac{1}{3m_{\Sigma_c}}(\gamma^\nu p^\alpha-\gamma^\alpha p^\nu)
+\frac{2}{3p_{\Sigma_c}^2}p_{\Sigma_c}^\nu p_{\Sigma_c}^\alpha]\\
&\quad\times u_{P_c\alpha}\frac{g_{P_c(4380)}g_{\Sigma_c^*\Lambda_c\pi}g_x}{(p_{\Sigma_c^*}^2-m_{\Sigma_c^*}^2)(p_{D}^2-m_D^2)(p_{\Lambda_c}^2-m_{\Lambda_c}^2)},\\
&\mathcal{M}^{P_c(4440)\to J/\psi p\pi\pi}\\
&=\int \frac{d^4p_D}{(2\pi)^4}(-g^{\alpha\rho}+\frac{p_{D^*}^\alpha p_{D^*}^\rho}{m_{D^*}^2})p_{\pi\rho}\\
&\quad\times\bar{u}_p\gamma_5\gamma_\mu \varepsilon_{J/\psi}^{*\mu}(\slashed{p}_{\Lambda_c}+m_{\Lambda_c})\gamma_5\gamma_\nu p_\pi^\nu(\slashed{p}_{\Sigma_c}+m_{\Sigma_c})\gamma_5\tilde{\gamma}_\alpha u_{P_c}\\
&\quad\times\frac{g_{P_c(4440)}g_{\Sigma_c^*\Lambda_c\pi}g_xg_{D^*D\pi}}{(p_{\Sigma_c}^2-m_{\Sigma_c}^2)(p_{D^*}^2-m_{D^*}^2)(p_{\Lambda_c}^2-m_{\Lambda_c}^2)(p_D^2-m_D^2)},\\
&\mathcal{M}^{P_c(4457)\to J/\psi p\pi\pi}\\
&=\int \frac{d^4p_D}{(2\pi)^4}  \bar{u}_p\gamma_5\gamma_\mu \varepsilon_{J/\psi}^{*\mu}(\slashed{p}_{\Lambda_c}+m_{\Lambda_c})\gamma_5\gamma_\nu p_\pi^\nu(\slashed{p}_{\Sigma_c}+m_{\Sigma_c})u_{P_c\alpha}\\
&\quad\times(-g^{\alpha\beta}+\frac{p_{D^*}^\alpha p_{D^*}^\beta}{m_{D^*}^2})p_{\pi\beta}\\
&\quad\times\frac{g_{P_c(4457)}g_{\Sigma_c^*\Lambda_c\pi}g_xg_{D^*D\pi}}{(p_{\Sigma_c}^2-m_{\Sigma_c}^2)(p_{D^*}^2-m_{D^*}^2)(p_{\Lambda_c}^2-m_{\Lambda_c}^2)(p_D^2-m_D^2)}.\\
\end{split}
\end{equation}
In the kinematic regions of the TS and BS, the internal particles are in the vicinity of their on-shell conditions. We further simplify the integrands by adopting the following eigenfunction relations for the spin-$1/2$, spin-$1$, and spin-$3/2$ particles can be simplified as $(\slashed{p}+m)\to 2m$, $(-g^{\mu\nu}+p^\mu p^\nu/p^2)\to \delta^{ij}$, and $(\slashed{p}+m)[-g^{\mu\nu}+\frac{1}{3}\gamma^\mu\gamma^\nu+\frac{1}{3m}(\gamma^\mu p^\nu-\gamma^\nu p^\mu)+\frac{2}{3m^2}p^\mu p^\nu]\to 2m[-g^{\mu\nu}+\frac{1}{3}\gamma^\mu \gamma^\nu+\frac{1}{3}(\gamma^\mu g^{\nu 0}-\gamma^\nu g^{\mu 0})+\frac{2}{3}g^{\mu 0}g^{\nu 0}]$. As a result, the numerators in the integrands do not contain the inner momentum and the loop integrals are actually convergent.

For the relatively broad states $\Sigma_c^{(*)}$, complex masses with the widths are adopted, while for $D^*$ we have ignored its width effects since they do not cause significant changes to the lineshapes of the $J/\psi p$ or $J/\psi p\pi$ invariant mass spectra.


{\it Results and discussions} \
Proceeding to the numerical calculations we first list in Table.~\ref{Tab1} the masses and widths adopted for these observed $P_c$ states~\cite{LHCb:2019kea,ParticleDataGroup:2022pth}. We stress that it is the TS and BS mechanisms  driving the enhancements expected at the $\Lambda_c\bar{D}$ and $\Lambda_c\bar{D}^*$ thresholds. These two threshold structures are insensitive to the uncertainties of the masses and widths of the $P_c$ states, and no need for us to re-derive these quantities here.

\begin{table}[htbp]
\caption{Masses (unit: MeV) adopted for the $P_c$ states. }
\label{Tab1}
\renewcommand\arraystretch{1.50}
\begin{tabular*}{86mm}{@{\extracolsep{\fill}}ccc}
\toprule[1.5pt]
\toprule[1pt]
$m_{P_c(4312)}=4312$ & $m_{P_c(4380)}=4380$ & $m_{P_c(4440)}=4440$ \\
$m_{P_c(4457)}=4457$ & $m_{\Sigma_c}=2453-i1.89/2$ & $m_{\Sigma_c^*}=2518-i15/2$ \\
$m_{\Lambda_c}=2286$&$m_p=938$&$m_{J/\psi}=3097$\\
$m_\pi=137$ & $m_D=1867$ & $m_{D^*}=2009$\\
\bottomrule[1pt]
\bottomrule[1.5pt]
\end{tabular*}
\end{table}

\begin{center}
\begin{figure}[htbp]
\includegraphics[width=8cm,keepaspectratio]{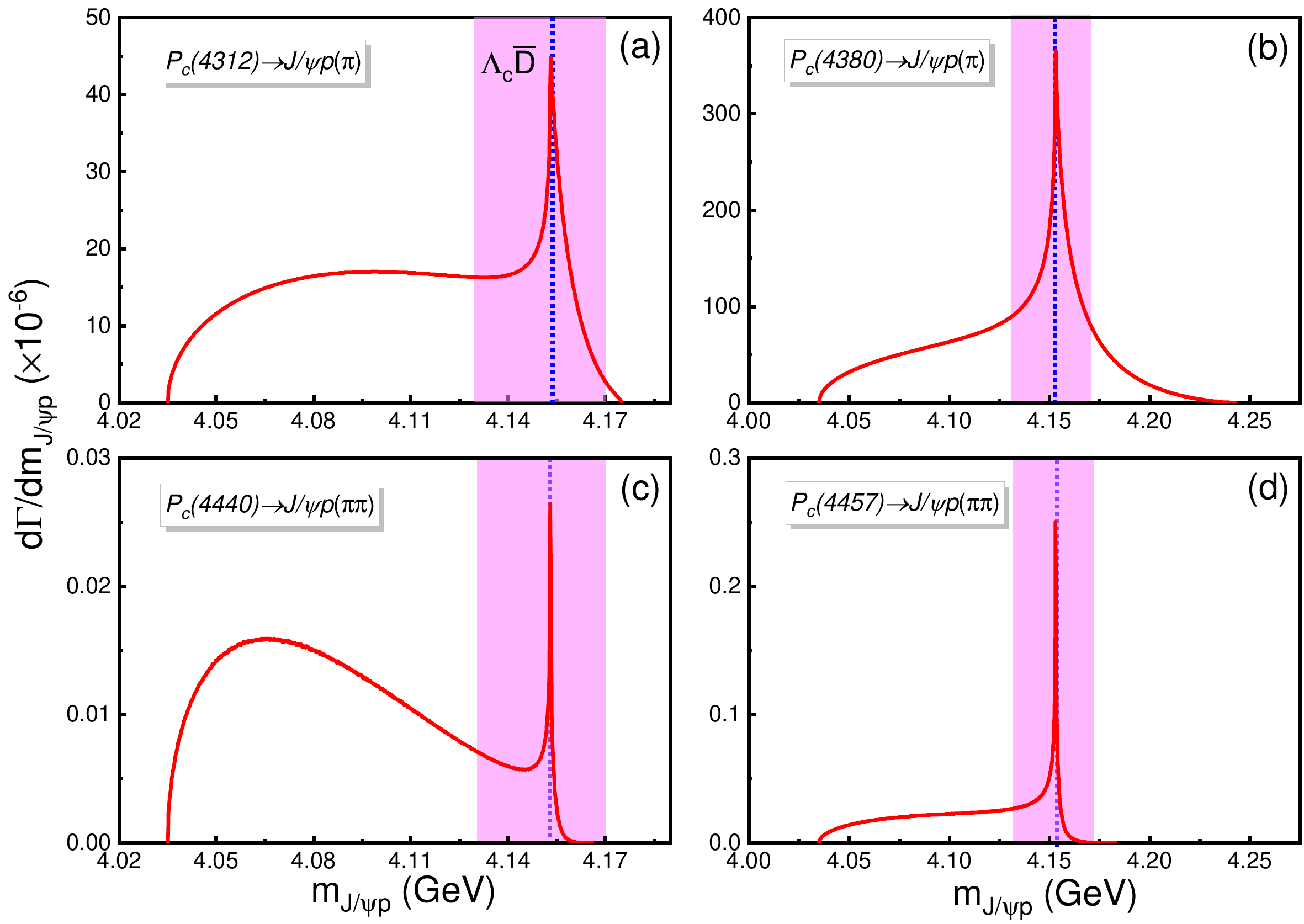}
\caption{The $J/\psi p$ invariant mass spectra (red solid lines) from different $P_c$ state decays. The dashed vertical lines indicate the $\Lambda_c\bar{D}$ threshold, and the purple bands denote the range of $4.13 {\rm GeV}<m_{J/\psi p}<4.17 {\rm GeV}$ the invariant mass spectra, which are collected in Fig.~\ref{f5}.}
\label{f3}
\end{figure}
\end{center}

For the feed-down processes, it is not necessary to have the initial states to be the pentaquark states. In principle, as far as the initial $\Sigma_c^{(*)}$ and $\bar{D}^{(*)}$ scatterings occur in an $S$-wave near threshold, they will feed down to contribute to the production of the threshold enhancements at the $\Lambda_c \bar{D}$ and $\Lambda_c\bar{D}^*$ thresholds in the $J/\psi p$ and $J/\psi p\pi$ invariant mass spectra, respectively, via the TS or BS mechanisms. If the initial states are genuine states, the enhancements will be strongly enhanced by the propagators when they are on-shell. Thus, even though we do the calculations with the $P_c$ states as the initial states, the feed-down mechanisms actually involve both possible genuine states and continuum contributions. This unique feature makes the semi-inclusive measurements of the $J/\psi p$ and $J/\psi p\pi$ spectra interesting and useful for clarifying not only the nature of the $\Lambda_c\bar{D}^{(*)}$ interactions, but also the role played by the TS and BS mechanisms.

In Fig.~\ref{f3} we plot the $J/\psi p$ invariant mass spectra for four observed $P_c$ states in their three-body or four-body decays into either $J/\psi p\pi$ or $J/\psi p\pi\pi$.  Clear and predominant peaks appear at the $\Lambda_c \bar{D}$ threshold due to the TS and BS mechanisms. Meanwhile, it shows that $P_c(4380)$ contributes the largest to the $J/\psi p$ spectrum. It is due to the relatively large phase space for $\Sigma_c^*\to \Lambda_c\pi$ within the TS kinematic region that strongly enhances the decay rate of $P_c(4380)\to J/\psi p\pi$. Such an effect actually will add additional partial decay widths to the $P_c(4380)$ decay~\footnote{We note that the TS also occurs in another situation in the three-body and four-body decays into $\Lambda_c\bar{D}\pi$ and $\Lambda_c\bar{D}\pi\pi$, respectively.  Different from Fig.~\ref{f1}, here the significant contributions from the $S$-wave coupling of $\bar{D}\pi$ into $\bar{D}_0(2400)$. These processes will also contribute to the total width of $P_c(4380)$ although it cannot be as large as 100 MeV suggested by the LHCb fits~\cite{LHCb:2019kea}. More detailed studies of such processes will be reported elsewhere.}.

\begin{center}
\begin{figure}[htbp]
\includegraphics[width=6cm,keepaspectratio]{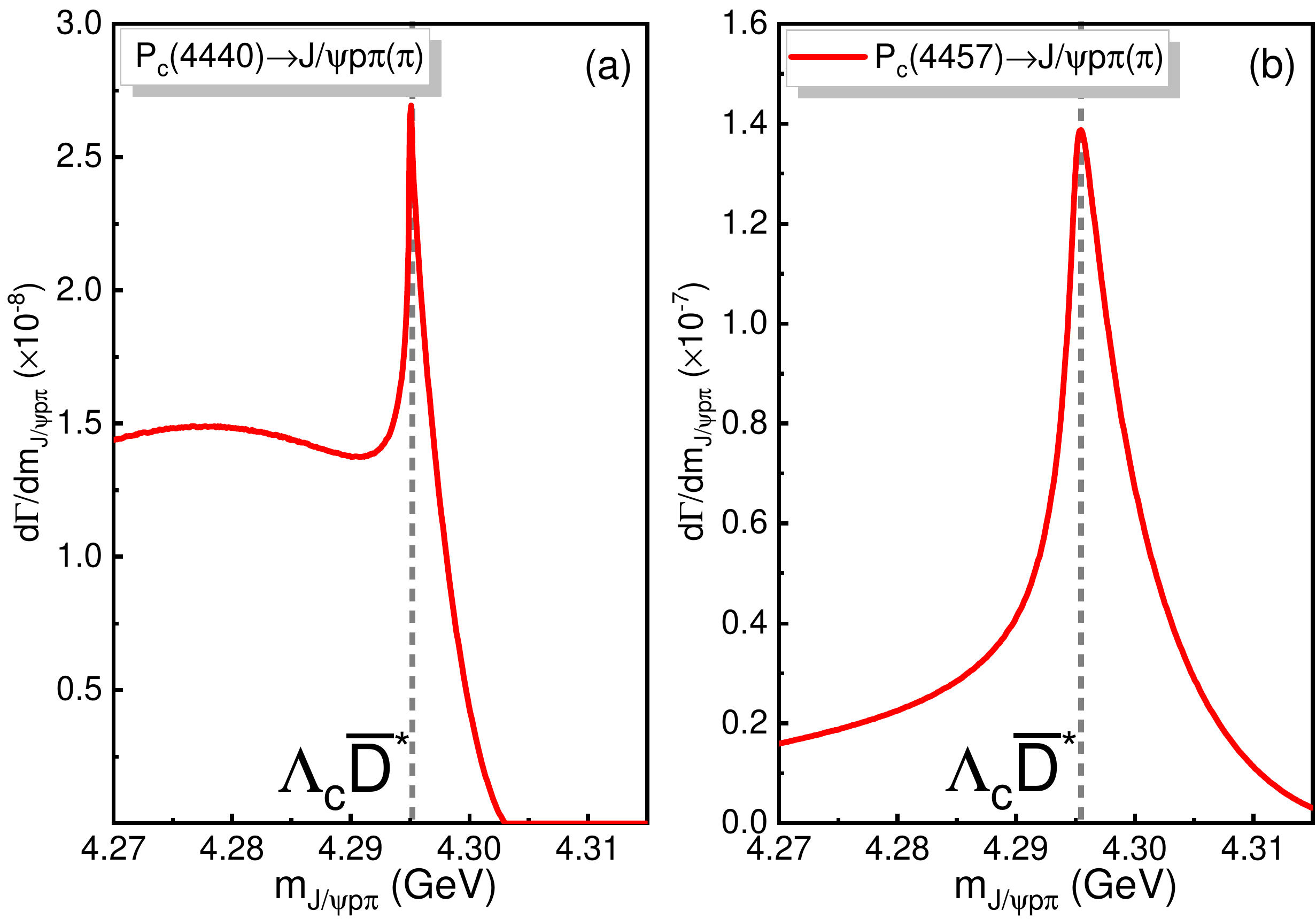}
\caption{The $J/\psi p\pi$ invariant mass spectra (red solid lines) from $P_c(4440)$ (left) and $P_c(4457)$ (right) decays. The $\Lambda_c\bar{D}^*$ threshold is indicated by the dashed vertical lines.}
\label{f4}
\end{figure}
\end{center}

The semi-inclusive $J/\psi p$ spectrum should contain contributions from the exclusive decay channels such as the $P_c$  discovery channel $\Lambda_b\to K^- J/\psi p$~\cite{LHCb:2015yax,LHCb:2019kea}. As shown in Ref.~\cite{Liu:2015fea}, the threshold of $\Lambda_c\bar{D}$ will appear as a two-body CUSP because the mass of $\Lambda_b$ is far above the thresholds of $\Lambda_c\bar{D}_s^*$ and $\Lambda_c\bar{D}_{s1}$ and the TS condition cannot be satisfied. This situation actually helps us clarify the role played by the TS and BS mechanisms since a narrow peak at the threshold of $\Lambda_c\bar{D}$ in the semi-inclusive $J/\psi p$ spectrum will indicate the leading TS and/or BS mechanisms instead of a CUSP effect~\cite{Guo:2014iya}.

The feed-down process via the BS mechanism also gives access to the $\Lambda_c\bar{D}^*$ threshold which can manifest as a narrow threshold enhancement as shown by the invariant mass spectrum of $J/\psi p\pi$ in Fig.~\ref{f4}. We stress that although such a threshold enhancement is illustrated by the four-body decays of $P_c(4440)$ and $P_c(4457)$, any processes involving the near-threshold $\Sigma_c^{(*)}\bar{D}^*$ $S$-wave scatterings may feed down to the $J/\psi p\pi\pi$ channel and produce the $\Lambda_c\bar{D}^*$ threshold enhancement. This phenomenon can also serve as a unique probe for the role played by the BS mechanism and can be searched in experiment.

In Fig.~\ref{f1} (c) in addition to the $\Lambda_c \bar{D}$ and $\Lambda_c\bar{D}^*$ in the invariant mass spectra of $J/\psi p$ and $J/\psi p\pi$, respectively, the $\Sigma_c^*\bar{D}$ threshold can also be accessed due to the BS mechanism. However, since the $\Sigma_c^*\bar{D}$ and $J/\psi p\pi$ thresholds are close to each other the $\Sigma_c^*\bar{D}$ threshold structure will be strongly suppressed.

It should be clarified that, although it is possible to observe the $\Lambda_c\bar{D}$ and $\Lambda_c\bar{D}^*$ threshold structures in exclusive processes, e.g. in $\Lambda_b\to K^- J/\psi p$ and $\Lambda_b\to K^- J/\psi p\pi$, the signal significance should be less than those of the $P_c$ states. The reason is that the relative $S$-wave interactions between $\Lambda_c$ and $\bar{D}^{(*)}$ are repulsive in the hadronic molecular picture. Without the pole structures, the signals accumulated from the $P_c$ decays due to the feed-down mechanism will be limited.

\begin{center}
\begin{figure}[htbp]
\includegraphics[width=6cm,keepaspectratio]{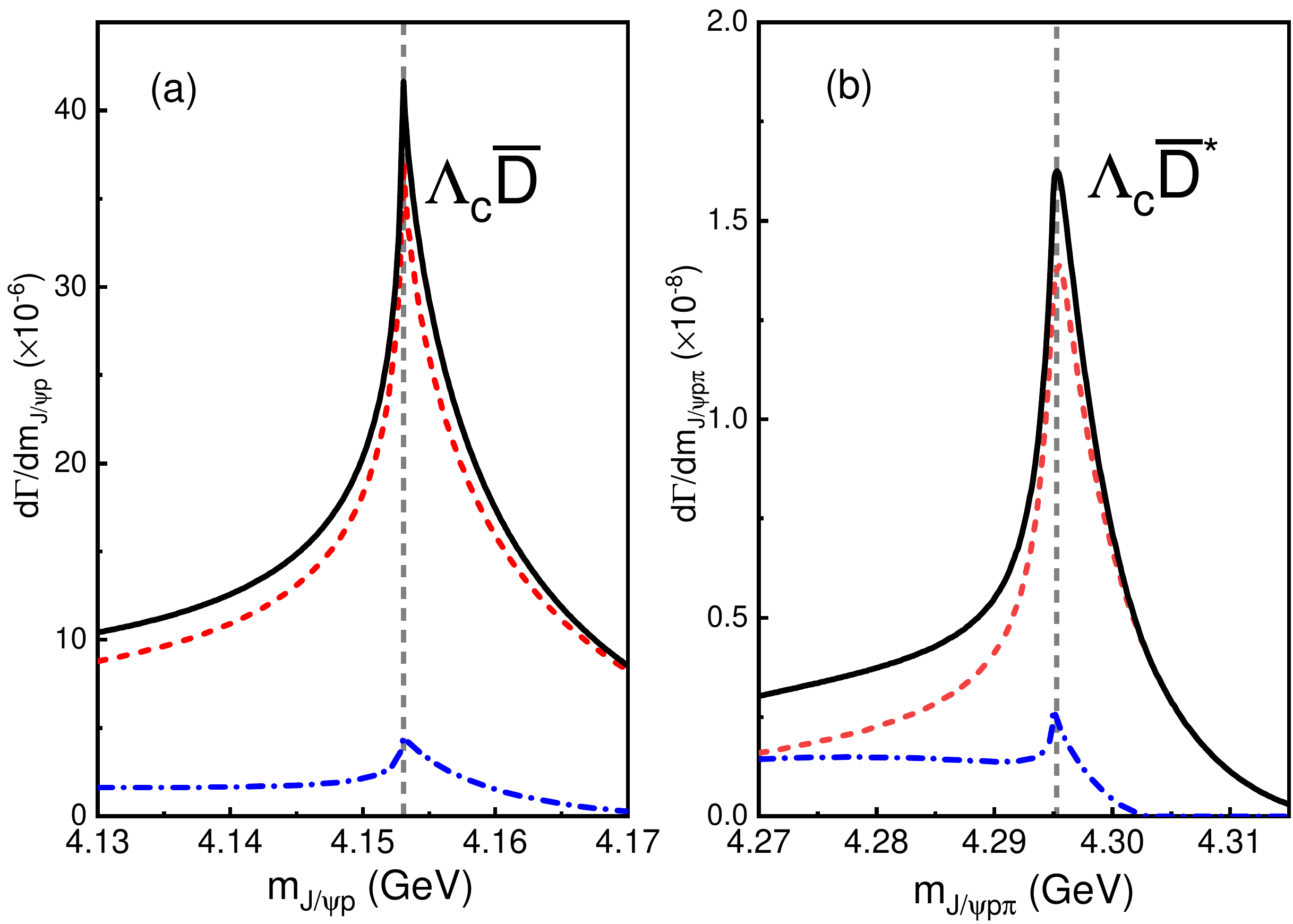}
\caption{The accumulated $J/\psi p$ (left) and $J/\psi p\pi$ (right) invariant mass spectra from all different $P_c$ states are shown by the black solid lines.  On the left panel the red dashed and blue dot dashed lines represent the results from $P_c(4380)$ and the sum of all the others, respectively. On the right panel the red dashed and blue dot dashed lines represent the results from $P_c(4457)$ and $P_c(4440)$, respectively.}
\label{f5}
\end{figure}
\end{center}

In order to demonstrate the feed-down effects we sum up the feed-down contributions from the $P_c$ decays and plot the $J/\psi p$ and $J/\psi p\pi$ spectra in Fig.~\ref{f5} (a) and  (b), respectively. We neglect the possible interferences at this moment. They may occur if the feed-down contributions are from the same exclusive process. One can see that as the lowest quark-rearranging threshold above the $J/\psi p$ channel, the TS and BS kinematics turn out to be unique. The signals originated from the TS and BS mechanisms will drop smoothly in the higher energy end. It means that these feed-down signals will stand predominantly on the incoherent background in the semi-inclusive measurement.

{\it Summary} \
We study the feed-down transitions of the observed $P_c$ states into $J/\psi p$ and $J/\psi p\pi$ via rescatterings, and predict that predominant narrow structures can be produced at the $\Lambda_c\bar{D}$ and $\Lambda_c\bar{D}^*$ thresholds due to the TS and BS mechanisms. Since the $S$-wave interactions between $\Lambda_c$ and $\bar{D}^{(*)}$ are repulsive in the hadronic molecule picture, a search for the $\Lambda_c\bar{D}$ and $\Lambda_c\bar{D}^*$ threshold enhancements can provide supportive evidences for the hadronic molecule interpretation for these $P_c$ states including  $P_c(4380)$. Meanwhile, it will also provide a unique way for pinning down the role played by the TS and BS mechanisms.  Since such feed-down transitions, in principle, can occur given the production of near-threshold $\Sigma_c^{(*)}\bar{D}^{(*)}$, we anticipate that the $\Lambda_c\bar{D}$ and $\Lambda_c\bar{D}^*$ threshold enhancements can be accessed in the semi-inclusive $J/\psi p$ and $J/\psi p\pi$ spectra in various processes, such as hadron-hadron collisions at LHCb, $e^+e^-$ annihilations at Belle-II, and $J/\psi$ photoproduction reactions at GlueX. Future experimental search for them are strongly recommended.


{\it Acknowledgement} \
This work is supported, in part, by the National Natural Science Foundation of China (Grant No. 12235018), DFG and NSFC funds to the Sino-German CRC 110 ``Symmetries and the Emergence of Structure in QCD" (NSFC Grant No. 12070131001, DFG Project-ID 196253076), National Key Basic Research Program of China under Contract No. 2020YFA0406300, and Strategic Priority Research Program of Chinese Academy of Sciences (Grant No. XDB34030302).

\end{document}